\documentstyle[epsfig,prb,aps,multicol]{revtex}
\begin{document}
\draft
\title{Peierls substitution in the energy dispersion of a hexagonal
       lattice}
\author{Gi-Yeong Oh}
\address{Department of Basic Science, Hankyong National University,}
\address{Kyonggi-do 456-749, Korea}
\date{\today}
\maketitle

\begin{abstract}
The method of the Peierls substitution in studying the magnetic subband
structure of a hexagonal lattice is re-examined. Several errors in the
formalism of a couple of recent papers are pointed out and rectified so
as to describe the effect of the magnetic field pertinently.
\end{abstract}

\begin{multicols}{2}

Recently, Fekete and Gumbs~\cite{GF99} reported on the study of the
energy spectrum of two-dimensional (2D) Bloch electrons subject to a
perpendicular sine-modulated magnetic field
\begin{equation}
 B_{z}(x,y)=B_{0}+B_{1}(x)
           =B_{0}+B_{1}\sin\left(2\pi x/T_{x}\right)          \label{1}
\end{equation}
by making the Peierls substitution $\vec{k}\rightarrow(\vec{p}+e
\vec{A})/\hbar$ in the first-order energy dispersion
$\varepsilon(\vec{k})$ obtained by the tight-binding method. The
content of Ref.~1 consists of two parts. The first part deals with the
formalism that describes the effect of magnetic modulation (i.e.,
$B_{1}\neq 0 $) on the energy spectrum of a {\it square lattice}, where
a correction of some errors in a paper~\cite{GF95} was given.
(Concerning this problem, it may be worthy of consulting Ref.~3.) The
second part deals with the energy spectrum of a {\it hexagonal lattice}
in the presence of magnetic modulation, where the authors attempted to
generalize the formalism presented in another paper~\cite{GF97} which
deals with the problem of the energy spectrum of the hexagonal lattice
under a uniform magnetic field. Through the studies, the authors of
Refs.~1,2, and 4 calculated energy eigenvalues, in the absence and/or
presence of magnetic modulation, by diagonalizing the effective
Hamiltonian obtained by the Peierls substitution. However, {\it as for
the hexagonal lattice}, the formalism and the numerical data presented
in Refs.~1 and 4 are flawed in several ways, whose origin lies in a
simple but fundamental mistake made at the early state of the Peierls
substitution. In this comment, we shall point out the mistake and the
relevant errors of Refs.~1 and 4, and derive an {\it exact} formalism
that enables to pertinently describe the effect of the magnetic field
given by (\ref{1}) on the energy spectrum of the hexagonal lattice. For
illustration purpose, we will present numerical data of the energy
eigenvalues obtained by diagonalizing the corrected matrix with and
without magnetic modulation.

To derive the Hamiltonian matrix that describes the effect of the
magnetic field given by (\ref{1}), we consider the energy dispersion of
a 2D hexagonal lattice given by
\begin{eqnarray}
 \varepsilon(\vec{k})=&&2\left\{t_{0}\cos(k_{x}a)
  +t_{+}\cos(k_{x}b+k_{y}c)\right.\nonumber\\
  &&\left.+t_{-}\cos(k_{x}b-k_{y}c)\right\}, \label{2}
\end{eqnarray}
where $a$ is the lattice constant, $b=a/2$, and $c=\sqrt{3}a/2$. Making
the Peierls substitution in (\ref{2}) yields the effective Hamiltonian
\begin{eqnarray}
 {\cal H}=&&t_{0}{\rm e}^{ip_{x}a/\hbar}
 +t_{+}{\rm e}^{i[p_{x}b+p_{y}c+eA_{y}(x)c]/\hbar}\nonumber\\
 &&+t_{-}{\rm e}^{i[p_{x}b-p_{y}c-eA_{y}(x)c]/\hbar}+{\rm c.c.},\label{3}
\end{eqnarray}
where
\begin{equation}
 A_{y}(x)=B_{0}x-(B_{1}T_{x}/2\pi)\cos(2\pi x/T_{x})          \label{4}
\end{equation}
is the $y$ component of the vector potential under the Landau gauge.

The four exponents of the form ${\rm e}^{Ap_{x}+Bp_{y}+f(x)}$ in
(\ref{3}) contain the operators $p_{x}$ and $x$ simultaneously in their
arguments, unlikely to the case of the square lattice, and thus we
should be very careful in manipulating them. Since $p_{y}$ commutes
with $p_{x}$ and $x$, we can easily decouple ${\rm
e}^{Ap_{x}+Bp_{y}+f(x)}$ as
\begin{equation}
 {\rm e}^{Ap_{x}+Bp_{y}+f(x)}={\rm e}^{Bp_{y}}{\rm e}^{Ap_{x}+f(x)}
  ={\rm e}^{Ap_{x}+f(x)}{\rm e}^{Bp_{y}}.                    \label{5}
\end{equation}
However, since $p_{x}$ does not commute with $x$, we cannot simply
decouple ${\rm e}^{Ap_{x}+f(x)}$ as
\begin{equation}
 {\rm e}^{Ap_{x}+f(x)}={\rm e}^{Ap_{x}}{\rm e}^{f(x)}~~{\rm or}~~
 {\rm e}^{f(x)}{\rm e}^{Ap_{x}}.                            \label{6}
\end{equation}
Instead, it should be written as
\begin{equation}
 {\rm e}^{Ap_{x}+f(x)}={\rm e}^{Ap_{x}}{\rm e}^{\int_{0}^{1}
  f(x+i\hbar A\lambda)d\lambda},                              \label{7}
\end{equation}
which is a generalization of the Weyl's formula and can be derived by
using the Baker-Hausdorff lemma \cite{Baym}. In Particular, for
$f(x)=Cx+D\cos(Kx)$, (\ref{7}) becomes
\begin{equation}
 {\rm e}^{Ap_{x}+Cx+D\cos(Kx)}={\rm e}^{Ap_{x}}
     {\rm e}^{C(x+i\hbar A/2)-iD\xi(x)/\hbar AK},          \label{8}
\end{equation}
where
\begin{equation}
 \xi(x)=\sin[K(x+i\hbar A)]-\sin(Kx).                       \label{9}
\end{equation}
Combining (\ref{5}) and (\ref{8}) with (\ref{3}) , we can write
\begin{eqnarray}
 {\cal H}=&&t_{0}\left({\rm e}^{ip_{x}a/\hbar}
  +{\rm e}^{-ip_{x}a/\hbar}\right) \nonumber\\
  +&&t_{+}{\rm e}^{-i\pi\alpha/2}\left({\rm e}^{ip_{y}c/\hbar}
  {\rm e}^{ip_{x}b/\hbar}{\rm e}^{i(\pi\alpha x'+\nu_{-})}
  \right. \nonumber\\
  &&+\left.{\rm e}^{-ip_{y}c/\hbar}{\rm e}^{-ip_{x}b/\hbar}
  {\rm e}^{-i(\pi\alpha x-\nu_{+})}\right) \nonumber\\
  +&&t_{-}{\rm e}^{i\pi\alpha/2}\left({\rm e}^{-ip_{y}c/\hbar}
  {\rm e}^{ip_{x}b/\hbar}{\rm e}^{-i(\pi\alpha
  x'+\nu_{-})}\right.\nonumber\\
  &&+\left.{\rm e}^{ip_{y}c/\hbar}{\rm e}^{-ip_{x}b/\hbar}
  {\rm e}^{i(\pi\alpha x'-\nu_{+})}\right),                \label{10}
\end{eqnarray}

\begin{figure}
\centerline{\epsfig{figure=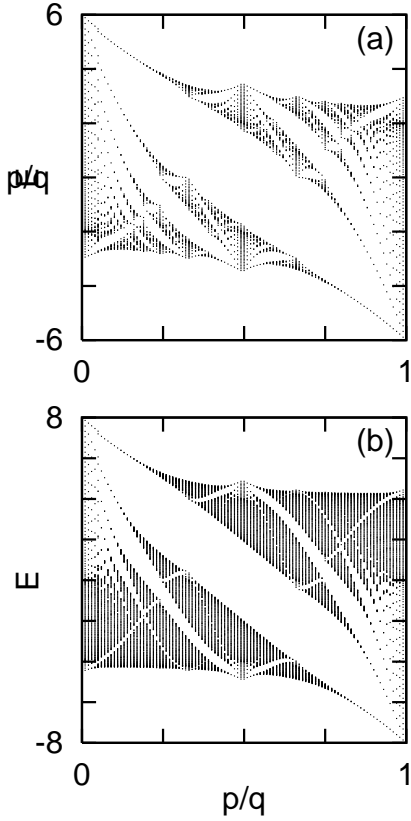,width=0.62\linewidth}}
\narrowtext
\caption{Energy eigenvalues versus $\alpha$ for $\gamma=0$ with (a)
  $(t_{0},t_{+},t_{-})=(1,1,1)$ and (b) $(t_{0},t_{+},t_{-})
  =(2,1,1)$. Calculations are performed for $\alpha=p/101$
  ($1\le p\le 100$) and  $(k_{x},k_{y})=(0,0)$.}           \label{fig1}
\end{figure}

\noindent where
\begin{equation}
 \nu_{\pm}(x')=\frac{\alpha\gamma}{\pi\beta^{2}}\left\{\sin[\pi\beta
 (x'\pm1)]-\sin(\pi\beta x')\right\}, \label{11}
\end{equation}
and
\begin{equation}
 \alpha=eB_{0}ac/2\pi\hbar,~\beta=a/T_{x},
 ~\gamma=B_{1}/B_{0},~x'=x/b,                       \label{12}
\end{equation}

Denoting the lattice points as $(m,n)=(x/b,y/c)$ and using the
translational property ${\rm
e}^{-i\vec{p}\cdot\vec{\xi}}|\vec{r}\rangle =|\vec{r}+\vec{\xi}
\rangle$, we can write the Schr\"{o}dinger equation
$\langle\vec{r}|{\cal H}|\psi \rangle=E\psi(\vec{r})$ as
\begin{eqnarray}
 E\psi_{m,n}=&&t_{0}\left(\psi_{m-2,n}+\psi_{m+2,n}\right)\nonumber\\
   +&&t_{+}\left({\rm e}^{-i\theta_{m-1}}\psi_{m-1,n-1}
   +{\rm e}^{i\theta_{m}}\psi_{m+1,n+1}\right) \nonumber\\
 +&&t_{-}\left({\rm e}^{i\theta_{m-1}}\psi_{m-1,n+1}
   +{\rm e}^{-i\theta_{m}}\psi_{m+1,n-1}\right),           \label{13}
\end{eqnarray}
where
\begin{eqnarray}
 \theta_{m}&&\left(\equiv \theta_{m}^{(0)} +\theta_{m}^{(1)}\right)
    =\pi\alpha\left(m+1/2\right) \nonumber\\
    &&-\frac{\alpha\gamma}{\pi\beta^{2}}
 \left\{\sin[(m+1)\pi\beta]-\sin(m\pi\beta)\right\}        \label{14}
\end{eqnarray}
is the magnetic phase factor. Note that $\theta_{m}$ plays a key role
in studying the magnetic subband structure within

\begin{figure}
\begin{center}
\centerline{\epsfig{figure=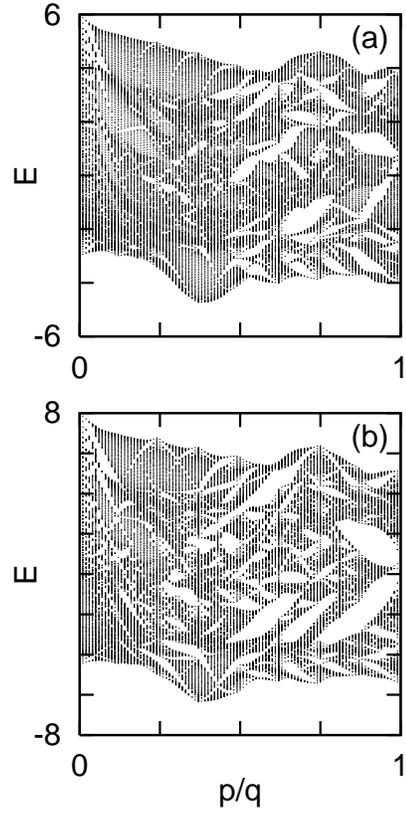,width=0.60\linewidth}}
\end{center}
\caption{The same as figure 1, except that $\gamma=2$ and
   $T_{x}=4$.}                                             \label{fig2}
\end{figure}

\noindent the tight-binding approximation since all the effects of the
applied magnetic field are held in it; $\theta_{m}^{(0)}$ reflects the
effect of $B_{0}$ while $\theta_{m}^{(1)}$ reflects the effect of
$B_{1}(x)$. Since the variable $y$ is cyclic under the Landau gauge, we
can write the wave function as $\psi(x,y)={\rm e}^{ik_{y}y}\psi(x)$,
which in turn enables us to write (\ref{13}) as
\begin{equation}
 E\psi_{m}=t_{0}\psi_{m-2}+\delta_{m-1}^{*}\psi_{m-1}
 +\delta_{m}\psi_{m+1}+t_{0}\psi_{m+2},                      \label{15}
\end{equation}
where
\begin{equation}
 \delta_{m}=t_{+}{\rm e}^{i\mu_{m}}+t_{-}{\rm e}^{-i\mu_{m}}       \label{16}
\end{equation}
with
\begin{equation}
 \mu_{m}=\theta_{m}+k_{y}c .                                 \label{17}
\end{equation}

Let us assume that $\alpha=p/q$ where $p$ and $q$ are coprime integers
and that $T_{x}(\geq 2)$ is an integer. Then, the period of
$\theta_{m}^{(0)}$ is $q$ ($2q$) for even (odd) $p$ and that of
$\theta_{m}^{(1)}$ is $2T_{x}$ if we set $a\equiv1$. Thus, the period
of $\theta_{m}$ is given by
\begin{equation}
 T=\left\{\begin{array}{ll}
 {\rm L.C.M.} (q, 2T_{x}) & {\rm for~even~} p\\
 {\rm L.C.M.} (2q, 2T_{x}) & {\rm for~odd~} p
 \end{array}\right. ,                                        \label{18}
\end{equation}
which leads to the relations $\mu_{m+T}=\mu_{m}$ and
$\delta_{m+T}=\delta_{m}$. Therefore, the Bloch condition along the $x$
direction can be expressed as $\psi_{m+T}={\rm e}^{ik_{x}Tb}\psi_{m}$,
and the characteristic  matrix that gives rise to energy eigenvalues in
presence of magnetic modulation can be written as
\begin{equation}
 {\bf A}=\left( \begin{array}{cccccccc}
  0 & \delta_{1} & t_{0} & 0 & \cdots & 0 &
    t_{0}{\rm e}^{-i\eta} & \delta_{T}^{*}{\rm e}^{-i\eta} \\
  \delta_{1}^{*} & 0 & \delta_{2} & t_{0} & \cdots & 0 &
    0 & t_{0}{\rm e}^{-i\eta} \\
  t_{0} & \delta_{2}^{*} & 0 & \delta_{3} & \cdots & 0 & 0 & 0\\
  0 & t_{0} & \delta_{3}^{*} & 0 & \cdots & 0 & 0 & 0 \\
  \vdots& \vdots& \vdots& \vdots& \ddots& \vdots& \vdots& \vdots \\
  0 & 0 & 0 & 0 & \cdots& 0 & \delta_{T-2} & t_{0} \\
  t_{0}{\rm e}^{i\eta} & 0 & 0 & 0 & \cdots & \delta_{T-2}^{*} &
    0 & \delta_{T-1} \\
  \delta_{T}{\rm e}^{i\eta} & t_{0}{\rm e}^{i\eta} & 0 & 0 & \cdots
    & t_{0} & \delta_{T-1}^{*} & 0
  \end{array}\right),                                        \label{19}
\end{equation}
where $\eta=k_{x}Tb$. Noted that, when $\gamma=0$, the period of
$\theta_{m}$, and thus the order of ${\bf A}$, is given by $T_{0}$
which equals to $q$ ($2q$) for even (odd) $p$.

We are now in position to discuss the formalism of Refs.~1 and 4. To
this end, we refer to (14) and (16) of Ref.~1 as ${\bf A}^{{\rm F}}$
and $\mu_{m}^{{\rm F}}$, and (6) of Ref.~4 as ${\bf A}^{{\rm G}}$,
respectively. The most fundamental mistake made by the authors of
Refs.~1 and 4 lies in that they employed (\ref{6}) instead of (\ref{7})
for developing the formalism. Since (\ref{6}) is meaningful only in the
$\hbar\rightarrow0$ limit, it is evident that the formalism in Refs.~1
and 4 does not suit in describing the quantum-mechanical behavior of
Bloch electrons in the presence of the magnetic field. To put it
concretely, $\mu_{m}^{{\rm F}}$ is an erroneous expression as a result
of the mistake; $\theta_{m}^{(0; {\rm F,G})}(\equiv 2\pi m\alpha)$ and
$\theta_{m}^{(1;{\rm F})}[\equiv-(\alpha\gamma /\beta)\cos(2\pi
m\beta)]$ should be replaced by $\theta_{m}^{(0)}$ and
$\theta_{m}^{(1)}$, respectively, in order to describe pertinently the
effect of the magnetic field given by (\ref{1}). Note that
$\mu_{m}^{{\rm F}}$ in turn has a crucial influence on the order of the
characteristic matrix. Since the period of $\theta_{m}^{(0;{\rm F,G})}$
is $q$ and that of $\theta_{m}^{(1;{\rm F})}$ is $T_{x}$, the period of
$\mu_{m}^{{\rm F}}$ becomes $T'=qT_{x}$, resulting in $\psi_{m+T'}={\rm
e}^{ik_{x}T'b} \psi_{m}$. Thus, the order of ${\bf A}^{{\rm F}}$ is
$T'$ for generic values of $q$ and $T_{x}$. However, as shown above,
the order of the corrected matrix ${\bf A}$ is not $T'$ but $T$.
Besides, the order of ${\bf A}$ under $\gamma=0$ is not $q$ but
$T_{0}$, unlike to the assertion of Ref.~4. Another indication that the
formalism of Refs.~1 and 4 is incorrect can be found in that ${\bf
A}^{{\rm F}}$ and ${\bf A}^{{\rm G}}$ are non-Hermitian, contrary to
the general idea that a matrix which gives rise to real eigenvalues
should be Hermitian. All these arguments in this paragraph indicate
that the relevant numerical data, i.e., figures~3 and 4 in Ref.~1 and
all the figures in Ref.~4, obtained by diagonalizing ${\bf A}^{{\rm
F}}$ and/or ${\bf A}^{{\rm G}}$ are not in the least the solutions of
the problem that the authors of Refs.~1 and 4 attempted to solve.

Figures~1(a) and 1(b) show plots of the energy eigenvalues versus
$\alpha$ in the absence of magnetic modulation (i.e., $\gamma=0$). We
can see that figure~1(a) is the same as figure~3 of Ref.~6 while quite
different from figure~1 of Ref.~4, which shows how terribly the energy
spectrum is changed when $\theta_{m}^{(0;F,G)}$ instead of
$\theta_{m}^{(0)}$ is employed in the calculation. Indeed, even though
the authors of Ref.~4 were aware that their result differs from that of
Ref.~6, they might have misunderstood that the two works
\cite{GF97,CW79} deal with exactly the same problem. Comparison of
figure~1(b) with figure~3 of Ref.~4 also shows how sensitively the
energy spectrum is affected by $\theta_{m}^{(0)}$ in the presence of
the hopping anisotropy. Figures~2(a) and 2(b) show plots of the energy
eigenvalues versus $\alpha$ in the presence of the magnetic modulation,
where the same parameters as in Ref.~1 are chosen for comparison. The
plots show that introducing magnetic modulation leads to very
complicated subband structure; gap closing and subband broadening
observed in the case of a square lattice \cite{Oh99,Oh96} can also be
seen. Comparison of figures~2(a) and 2(b) with figures~3 and 4 of
Ref.~1 also shows the importance of the correct choice of
$\theta_{m}^{(1)}$. For instance, the straight lines of $E=-2$ and/or
$E=-4$ along the $\alpha$ axis in some plots of Refs.~1 and 4 do not
appear if we use correct form of the magnetic phase factor [i.e.,
(\ref{14})] in the calculation.

Before ending up this comment, we would like to mention two remarks.
One is that, since we focused our attention in this comment on the
formalism of Refs.~1 and 4,  detailed description of the effect of
magnetic modulation on the energy spectrum of the hexagonal lattice is
still lacking and thus further study on this problem is required for
better understanding. The other is that (\ref{13}) and thus (\ref{19})
can also be obtained without difficulty if we starts directly from the
tight-binding Hamiltonian given by
\begin{equation}
 {\cal H}=\sum_{ij}t_{ij}
          {\rm e}^{i\theta_{ij}}|i\rangle\langle j|,       \label{20}
\end{equation}
where $\theta_{ij}=(e/\hbar)\int_{i}^{j}\vec{A}\cdot d\vec{l}$. Indeed,
it can be easily checked that
\begin{eqnarray}
 &&\theta_{m,n;m\pm2,n}=0,~ \theta_{m,n;m+1,n\pm1}
 =\pm\theta_{m},\nonumber\\
 &&\theta_{m,n;m-1,n\pm1}=\pm\theta_{m-1},               \label{21}
\end{eqnarray}
and that combining (\ref{21}) with (\ref{20}) leads to (\ref{13}),
which confirms the correctness of the formalism derived in this
comment.


\end{multicols}
\end{document}